\documentclass[11pt]{article}
\usepackage[margin=1in]{geometry}
\usepackage{amsmath,amssymb}
\usepackage[T1]{fontenc}
\usepackage[utf8]{inputenc}
\usepackage{lmodern}
\usepackage{graphicx}
\usepackage{microtype}

\title{How the Quantum Sorites Phenomenon Strengthens Bell's Argument\\ --- and How a Random-Matrix Collapse Dynamics Answers It}
\author{Malcolm R. Forster\\ \normalsize University of Wisconsin--Madison}
\date{July 2026}

\begin{document}
\maketitle

\begin{abstract}
\noindent Bell proved that no theory of pre-existing local
values can reproduce the predictions of quantum mechanics, but his proof
leaves the culprit ambiguous: one may reject either of two independence
conditions, Outcome Independence or Parameter Independence, and most
commentators have found Outcome Independence the safer sacrifice. The
first part of this paper presents, in a form adapted to spin-1/2
particles in the singlet state, an argument (Forster 2014, building on
Colbeck and Renner 2011) that removes the ambiguity: using a chained
family of experiments in which quantum mechanics predicts an extreme
pattern of correlations --- the \emph{quantum Sorites} phenomenon --- a
contradiction is derived without ever assuming Outcome Independence.
Under the resulting theorem, anyone who holds that hidden variables
could improve on the quantum probabilities must give up Parameter
Independence itself. That looks like a heavy price, because Parameter
Independence appears to be protected twice over: rejecting it seems to
put superluminal influences into spacetime, and its statistical shadow
--- the No-Signaling condition --- is experimentally beyond
reproach.\footnote{The 2014 paper was accepted by the Philosophy of
  Science Association program committee for presentation at the PSA 2014
  meetings, but was not accepted for publication in \emph{Philosophy of
  Science}; it has been available in the physics archives since. Part 1
  of the present paper supersedes it; Part 2 is new.} The second part of
the paper shows that the price is payable. In the random-matrix collapse
dynamics proposed by Kryukov, measurement is a random walk of the
quantum state, and the hidden variable is not a stock of values fixed at
the source but the random stream that drives the walk --- like the
stored random numbers of a computer simulation, with the measurement
settings playing the role of seeds. In that framework Parameter
Independence is false while Outcome Independence and No-Signaling are
both true, and one can say exactly how the Sorites argument is blocked,
why the violation involves no process propagating in spacetime, and why
the influence of one wing's setting on the other wing's outcome ---
demonstrated here in a simulation, down to angle differences of
\(10^{-10}\) degrees --- can never be used to send a message.
\end{abstract}

\section{Introduction}

Bell (1964) derived a contradiction from four assumptions: that the
probabilistic predictions of quantum mechanics are correct in a family
of two-wing experiments, and that the outcomes are screened off from one
another by a hidden variable \(\lambda\) in specific senses that the
later literature disentangled into separate conditions.\footnote{The
  disentangling is usually credited to Jarrett (1984), but it is older.
  Clauser and Horne (1974) already construct their factorizability
  condition as the conjunction of the two independences, though without
  naming them; Suppes and Zanotti (1976) isolate the
  outcome-independence component and prove that, joined to the singlet's
  strict correlations, it forces determinism; and van Fraassen (1982)
  separates and names cognate conditions (his Hidden Locality and
  Causality) two years before Jarrett's completeness/locality analysis.
  The now-standard names, used here, are Shimony's (1986).} Two of those
senses matter here. \emph{Parameter Independence} (PI) says that, once
\(\lambda\) is fixed, the probability of an outcome on one wing does not
depend on the \emph{setting} chosen on the other wing. \emph{Outcome
Independence} (OI) says that, once \(\lambda\) and both settings are
fixed, the probability of an outcome on one wing does not depend on the
\emph{outcome} obtained on the other. Bell's contradiction shows that at
least one premise is false; since the quantum predictions are
overwhelmingly confirmed, a hidden-variable theorist must reject PI or
OI (or the autonomy of \(\lambda\) from the settings, of which more
below). And for fifty years the standard verdict has been that OI is the
safe thing to reject: a violation of OI has seemed the metaphysically
tamer failure --- a holism of the outcomes, with no agent's hand in it
--- while a violation of PI looks like an \emph{action}: a setting is an
agent's free choice, and if my choice makes a difference to your outcome
at spacelike separation, it seems something must have travelled faster
than light (Jarrett 1984; Butterfield 1989; Ghirardi 2010). (Where the
blame ultimately falls is a nice irony, previewed here and delivered in
section 11: in the framework of Part 2 it is OI that survives --- a
complete common cause screens the outcomes off from one another, exactly
as the Causal Markov Condition says it must --- and PI that fails.)

The first part of this paper presents an argument that this comfortable
verdict cannot stand. The argument descends from a theorem of Colbeck
and Renner (2011), which I recast in 2014 in a form that makes its
logical relation to Bell's theorem explicit; I called the result the
Stronger Theorem, and the extreme quantum prediction that powers it the
\emph{quantum Sorites} phenomenon. The Stronger Theorem derives a
contradiction \emph{without assuming Outcome Independence at all}. Its
premises are: the quantum predictions in a chained family of
experiments; the standard autonomy (``free choice'') condition on
\(\lambda\); Parameter Independence; and the assumption --- constitutive
of the hidden-variable programme --- that conditioning on \(\lambda\)
improves the quantum probabilities non-trivially and robustly. Since the
last assumption is the point of hidden variables, and the autonomy
condition is one nobody wishes to deny, the Stronger Theorem aims the
arrow of \emph{modus tollens} where Bell's theorem could not: at
Parameter Independence itself. Sections 2--7 present the argument,
adapted throughout to the case Bell himself began with --- two spin-1/2
particles in the singlet state --- and section 6 states exactly what
must be added to the 2014 presentation to make the inconsistency
deductively rigorous rather than rigorous-in-the-limit.

That leaves a puzzle, which section 8 sharpens. Parameter Independence
seems true. It seems true for a metaphysical reason: if all physical
processes run their course in four-dimensional spacetime, then a
violation of PI at spacelike separation is a superluminal process, and
special relativity forbids those. And it seems true for an inductive
reason: averaging PI over \(\lambda\) yields the No-Signaling condition
--- the statistical independence of each wing's outcomes from the other
wing's settings --- and No-Signaling is as well confirmed as anything in
physics --- indeed it is itself one of the predictions of quantum
mechanics, a fact about what Part 1 will call the surface probabilities,
so it inherits the entire experimental case for the theory. No-Signaling
does not entail PI; but its truth is evidence for PI in the way the
truth of a consequence is always evidence for its source.

The second part of the paper (sections 9--12) dissolves the puzzle by
exhibiting a concrete collapse dynamics in which the Stronger Theorem's
verdict is simply accepted: Parameter Independence is false,
No-Signaling is true, and one can see in detail how both halves come
about. The framework is the random-matrix measurement dynamics developed
by Kryukov (2020; 2025; 2026; 2026a; 2026b), in which measurement is a
stochastic-but-unitary random walk of the quantum state, driven by
random Hamiltonian matrices, that terminates on a classical outcome
state with Born-rule probabilities. The key move is a reinterpretation
of \(\lambda\) that the theorems of Part 1 permit but the literature
rarely considers: \(\lambda\) is not a stock of outcome values (or
dispositions) attached to the particles at the source; it is the
\emph{random stream} that drives the measurement walk --- in
computational terms, the long list of stored pseudo-random numbers a
computer uses to simulate chance, with the experimenters' settings
playing the role of seeds. Under that reading, the outcome of Bob's
measurement is a deterministic function of his stream \(\lambda_b\), his
setting \(b\), and the state presented to his device; and because the
walk's landing depends sensitively on \(b\) at fixed \(\lambda_b\),
Parameter Independence fails --- demonstrably, and in simulation
spectacularly, with outcomes flipping under setting changes of less than
\(10^{-10}\) degrees. Yet averaging over the stream restores the quantum
statistics exactly, so No-Signaling holds as a theorem of the model. The
influence of \(b\) on the distant outcome is real; it propagates through
the geometry of the joint state space rather than through spacetime; and
its chaotic character is precisely what makes it useless for sending
messages.

The moral of the whole is best said up front. The quantum Sorites
argument forces a choice that Bell's theorem left open, and the choice
falls on Parameter Independence. What Part 2 adds is that this is a
bullet a serious physical theory can bite --- indeed, in the
random-matrix framework, biting it is painless: the two reasons for
believing PI dissolve on inspection, because the framework denies that
the collapse process runs its course in spacetime (dissolving the
relativistic worry) while entailing No-Signaling outright (respecting
the evidence that supported PI in the first place).

\section{Part 1: The Quantum Sorites
Argument}

\section{The Quantum Sorites
Phenomenon}

The Bell argument is a \emph{reductio ad absurdum} --- the form of
argument that refutes a set of assumptions by showing that, taken
together, they entail something absurd. If the absurdity really follows,
then at least one of the assumptions is false; the argument itself does
not say which, and the interesting work always lies in deciding where
the blame falls. Bell's reductio derives a contradiction from the main
assumption that the probabilistic predictions of quantum mechanics are
correct in a particular family of experiments, together with auxiliary
assumptions about a hypothetically postulated variable \(\lambda\). The
important difference between the Stronger Theorem and Bell's theorem
lies in \emph{which} experiments are used. The Stronger Theorem uses a
chained family in which quantum mechanics predicts an extreme kind of
phenomenon --- the quantum Sorites phenomenon. It is the strength of
this predicted phenomenon that makes it possible to derive a
contradiction from substantially weaker assumptions about \(\lambda\).
There is a sense in which the total strength of the premises is the same
in both theorems; what differs is how the strength is divided between
the quantum predictions and the \(\lambda\)-assumptions. Since the
quantum predictions are confirmed to extraordinary accuracy, nobody
wants to resolve the contradiction by blaming them; the arrow of
\emph{modus tollens} points at the \(\lambda\)-assumptions, and the
weaker those are, the more damaging the conclusion.

Here is the phenomenon, for the singlet state. A source produces pairs
of spin-1/2 particles --- silver atoms will do --- in the singlet state

\[|\Psi_{s}\rangle \;=\; \frac{1}{\sqrt{2}}\, |z\text{-up}\rangle_{A}\, |z\text{-down}\rangle_{B} \;-\; \frac{1}{\sqrt{2}}\, |z\text{-down}\rangle_{A}\, |z\text{-up}\rangle_{B},\]

one particle flying to Alice's laboratory on the left, the other to
Bob's on the right. Each laboratory contains a Stern--Gerlach magnet
whose orientation the local experimenter controls: Alice's magnet is set
at angle \(a\), Bob's at angle \(b\), both measured from a shared
reference direction in a fixed plane, and each ranging from
\(0^{\circ}\) to \(180^{\circ}\). Neither experimenter controls the
outcome. Alice's particle is deflected toward one of two detector
regions, and her record is the value of a variable \(X\): she writes
\(X = 1\) if her particle lands in her \emph{up} region and \(X = 0\) if
it lands in her \emph{down} region. For reasons that will be plain in a
moment, Bob uses the opposite bookkeeping: he writes \(Y = 1\) for
\emph{down} and \(Y = 0\) for \emph{up}. (The relabeling is pure
bookkeeping; it recodes the singlet's perfect anti-correlations as
perfect correlations of the recorded bits, which lets every formula
below be read as ``the records agree'' or ``the records differ.'') We
assume every pair produces a record on each wing.

Two elementary facts about the singlet are all we need. (1) When the
angle between the two measurement directions is small, the recorded bits
are very probably equal: \(P(X = Y) \approx 1\). (2) When the directions
differ by \(180^{\circ}\), the records are perfectly anti-correlated:
\(P(X \neq Y) = 1\). Quantitatively, quantum mechanics puts the
probability of a mismatch at

\[P(X \neq Y \mid A = a, B = b) \;=\; \sin^{2}\!\left(\frac{\Delta}{2}\right), \qquad \Delta = a - b,\]

which for small \(\Delta\) is approximately \(\Delta^{2}/4\) ---
\emph{quadratically} small (Figure 2). This flatness near \(\Delta = 0\)
is the engine of everything that follows.

Now chain the settings (Figure 1). Fix a small angle
\(\varepsilon = 180^{\circ}/N\), and let Bob's settings be the even
multiples \(b = 0, 2\varepsilon, 4\varepsilon, \ldots\) while Alice's
are the odd multiples
\(a = \varepsilon, 3\varepsilon, \ldots, 180^{\circ} - \varepsilon\).
Each of Bob's angles is adjacent (a difference of \(\varepsilon\)) to
two of Alice's angles, and vice versa, giving a chain of experiments in
which each adjacent pair of settings yields records that agree with
probability \(1 - \sin^{2}(\varepsilon/2)\), which tends to \(1\) as the
chain is refined. Finally the two ends of the chain --- Alice at
\(180^{\circ} - \varepsilon\) and Bob at \(0\) --- are linked by an
experiment in which the settings differ by (nearly) \(180^{\circ}\), so
the records there \emph{disagree} with probability tending to \(1\).

\begin{figure}[t]
\centering
\includegraphics[width=\linewidth]{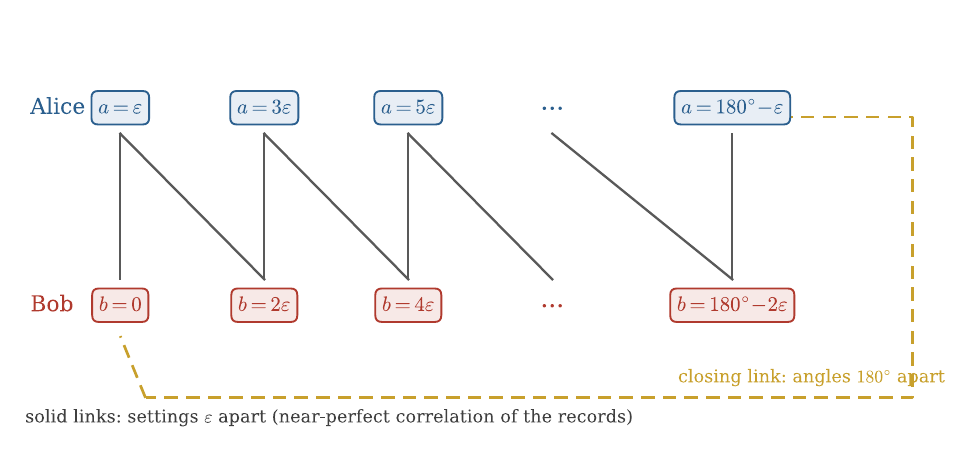}
\caption{The quantum Sorites phenomenon arises from a chained family of measurement settings on singlet pairs. Solid links join adjacent settings (\(\varepsilon\) apart), where the recorded bits agree with probability approaching one; the dashed closing link joins the two ends of the chain, where the settings differ by \(180^{\circ}\) and the bits disagree with certainty.}
\end{figure}

The Sorites character of the phenomenon is now visible, and it gives the
phenomenon its name. In the ancient paradox, adding a single grain of
sand makes no noticeable difference at any step --- one grain is not the
difference between a non-heap and a heap --- and yet the end of the
process differs utterly from the beginning: grain by unnoticeable grain,
a heap has appeared. Likewise here. Each small change of measurement
angle makes no noticeable difference to the correlation: every link of
the chain says, with probability as close to one as we please, \emph{the
records at these two adjacent settings are equal}. Following the links
from one end of the chain to the other, we may reasonably expect the two
end settings to be perfectly correlated as well. Instead, the closing
link says the two end records are perfectly \emph{anti-correlated} ---
as far from the expected perfect correlation as it is possible to be.
The single grain has become a heap. The total probability of \emph{any}
mismatch along the whole chain is at most
\(N \sin^{2}(\varepsilon/2) \approx \pi^{2}/4N\), which tends to zero as
\(N\) grows: quantum mechanics predicts that, with probability
approaching one, \emph{every} adjacent pair of records agrees while the
end pair disagrees.\footnote{The idea of chaining Bell experiments goes
  back at least to Braunstein and Caves (1990); the observation that
  perfect correlations alone strengthen Bell's argument was developed by
  Graßhoff, Portmann and Wüthrich (2005) and generalized in Hofer-Szabó,
  Rédei and Szabó (2013), though in those treatments Outcome
  Independence is built into the common-cause framework itself. The
  limit theorem that the chained probabilities cannot be improved is due
  to Colbeck and Renner (2011, supplement).} It will be useful to have a
name for the idealized limit. Following the terminology of Forster
(1986), call these the \textbf{Extreme Non-classical probabilities}:
mismatch probability exactly \(0\) for adjacent settings and exactly
\(1\) for the closing link (Figure 2). Let it be said plainly at first
mention: the Extreme Non-classical probabilities are false --- the true
mismatch probability at adjacent settings is quadratically small, not
zero --- and they will appear in this paper only as a pedagogical device
for exhibiting the logical skeleton of the argument, never as a premise
of a theorem. The theorem itself, in section 6, uses only the true
quantum probabilities.

\begin{figure}[t]
\centering
\includegraphics[width=\linewidth]{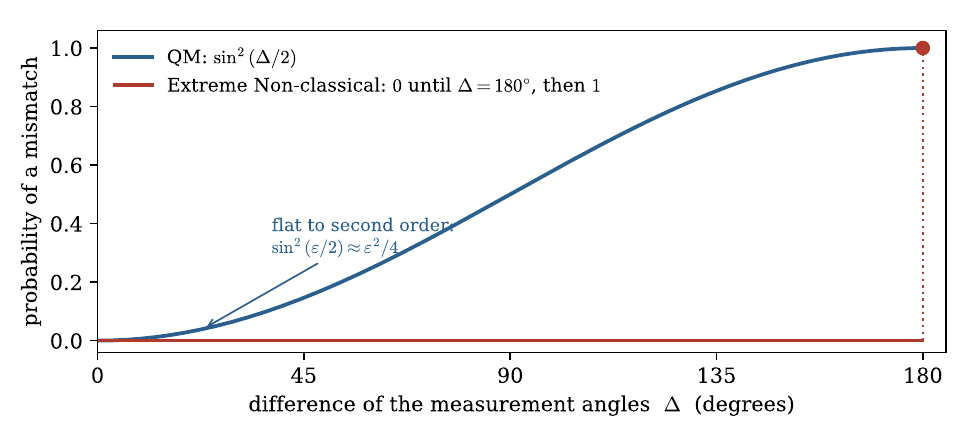}
\caption{The quantum probability of a mismatch between the recorded bits as a function of the difference in measurement angles, \(\sin^{2}(\Delta/2)\), and its Extreme Non-classical idealization: no mismatches until the settings are fully opposed, then certain mismatch. The quadratic flatness at small angles is what allows a chain of \(N\) near-certainties to remain a near-certainty jointly.}
\end{figure}

The fact that pairs of particles can orchestrate their behavior in this
way is truly remarkable, given that the events in Alice's laboratory may
be spacelike separated from the events in Bob's. The analogy of the next
section is designed to show \emph{why} the phenomenon implies some kind
of non-local dependence --- and to show it without writing a single
probability formula.

\section{The Paint-Chip Example}

Consider an ESP experiment built from the most familiar Sorites series
there is: a paint-store sample strip, shading in a thousand tiny steps
from pure blue at one end to pure green at the other. Neighbouring chips
are indistinguishable in colour --- hold any two side by side and no one
can tell them apart --- yet the two ends differ as plainly as sky from
grass. That is the ancient paradox in commercial form, and the reader
has seen the quantum version already: Figure 2's mismatch curve is flat
for neighbouring settings and maximal for opposed ones.

Alice and Bob are taken into separate rooms. Each is handed one chip
from the strip and asked a single question --- ``Blue?'' --- to be
answered `yes' or `no'; the answers are the recorded outcomes. The rules
of the game, announced in advance, mirror the chained experiments: on
almost every run the two chips handed out are immediate neighbours on
the strip, and the pair \emph{wins} if Alice and Bob give the
\emph{same} answer (their chips are indistinguishable, so they should
agree). Alice and Bob settle on the simplest plan in the waiting room:
Bob will answer `yes' --- \emph{blue} --- no matter what he is handed,
and Alice will answer `yes' too, since her chip is (almost always) next
to his. This wins every ordinary run. On the occasional exceptional run,
however, Alice is handed the greenest chip and Bob the bluest, and the
pair wins only if their answers are \emph{opposite}.

But what should Alice do when the greenest chip lands in her hand? If
Bob has been handed its neighbour --- the second-greenest chip --- this
is an ordinary run: he will say `yes', so she must say `yes'. If Bob has
been handed the bluest chip, this is the exceptional run: he will say
`yes', so she must say `no'. Her own chip looks exactly the same in both
cases. What she needs to know is \emph{which chip Bob was handed} ---
his setting. She asks the experimenter, who doesn't know. So she
guesses. And she gets it right. Repeatedly. Every run on which she says
`yes' turns out to be one where Bob holds the second-greenest chip;
every run on which she says `no' turns out to be the exceptional one. If
every ordinary channel of communication has been excluded, we seem
forced to say that she somehow tacitly knows what Bob was handed.

Note carefully what would \emph{not} help her: being told Bob's
\emph{answer}. Bob says `yes' on every run; his answer carries no
information whatever about his chip. What Alice needs is Bob's
\emph{setting}. In other words, the puzzle has nothing to do with
Outcome Independence --- no amount of outcome-to-outcome correlation
explains her success. If Alice reliably chooses correctly, there is a
dependence of her behavior on Bob's setting: a violation of Parameter
Independence, under any assignment of hidden helps and hints one likes.
Nor does it help to let Alice and Bob plan differently in the waiting
room: whatever conditional strategy they adopt, one of them ends up
needing the other's setting in some problematic case. The behavior of
singlet pairs in the chained experiments is puzzling for exactly this
reason, and the theorem of sections 5--6 is essentially a proof that
there is no way around the problem.

\section{The Colbeck--Renner Lemma and the Surface
Argument}

What I call the Colbeck--Renner Lemma is a simple fact of probability
theory; it is the mathematical device that lets the Stronger Theorem
bypass Outcome Independence.

\textbf{Lemma (Colbeck--Renner).} Let \(X\) and \(Y\) be two-valued
variables. (a) If \(P(X = Y) = 1\) then \(P(X = 1) = P(Y = 1)\). (b) If
\(P(X \neq Y) = 1\) then \(P(X = 1) = 1 - P(Y = 1)\). More generally,
without any assumption: (a\('\))
\(\lvert\, P(X = 1) - P(Y = 1) \,\rvert \leq P(X \neq Y)\), and (b\('\))
\(\lvert\, P(X = 1) - \bigl(1 - P(Y = 1)\bigr) \,\rvert \leq P(X = Y)\).

\emph{Proof.} For (a\('\)):
\(P(X{=}1) - P(Y{=}1) = P(X{=}1, Y{=}0) - P(X{=}0, Y{=}1)\), and each
term on the right is at least \(0\) and at most \(P(X \neq Y)\). For
(b\('\)):
\(P(X{=}1) - \bigl(1 - P(Y{=}1)\bigr) = P(X{=}1) + P(Y{=}1) - 1 = P(X{=}1, Y{=}1) - P(X{=}0, Y{=}0)\),
and each term on the right is at least \(0\) and at most \(P(X = Y)\).
Part (a) is (a\('\)) in the case \(P(X \neq Y) = 0\), and part (b) is
(b\('\)) in the case \(P(X = Y) = 0\). \(\square\)

The exact parts (a) and (b) drive the idealized argument of this section
and the next; the error parts (a\('\)) and (b\('\)) will be the engine
of the real theorem in section 6, where the correlations are nearly but
not exactly perfect.

The Lemma needs nothing beyond the axioms of probability, and it applies
equally to conditional probabilities, provided all are conditional on
the same proposition and well defined. Its significance is that it
converts a fact about the \emph{joint} distribution of the two wings (a
perfect correlation) into a relation between the \emph{marginal}
distributions of each wing separately --- and marginals are what
No-Signaling and Parameter Independence speak about. That is the trick
that will make Outcome Independence unnecessary: where Bell needs OI to
factorize joint probabilities, the Sorites argument only ever
manipulates marginals, which the perfect correlations tie to one another
via the Lemma.

Two well-confirmed assumptions govern the \emph{surface} probabilities
--- van Fraassen's (1982) term for the probabilities conditional on
settings alone, with no \(\lambda\) in sight. Both are confirmed
directly by experiment; No-Signaling is, in addition, itself a
prediction of quantum mechanics --- a Born-rule fact about the surface
probabilities --- and so is indirectly supported by the entire
experimental case for quantum theory.

\textbf{Weak Autonomy.} Let \(S\) be the set of setting pairs \((a, b)\)
that define the chain. For all \((a, b) \in S\):
\(P(A = a, B = b) > 0\).

This says only that every experiment in the chain can actually be run:
no conspiracy prevents any of the chained setting pairs from occurring.
(It is called \emph{weak} because it constrains only the settings, not
\(\lambda\).)

\textbf{No-Signaling.} For all \((a, b) \in S\):
\(P(X = 1 \mid A = a, B = b) = P(X = 1 \mid A = a)\) and
\(P(Y = 1 \mid A = a, B = b) = P(Y = 1 \mid B = b)\). (This is van
Fraassen's (1982) Surface Locality; No-Signaling is the standard name,
and the one used throughout this paper.)

No-Signaling says that each wing's outcome statistics depend only on
that wing's setting. It licenses unambiguous names for the marginals:
write \(p_{2j}\) for \(P(Y = 1 \mid B = 2j\varepsilon)\) --- Bob's
marginals, even subscripts --- and \(q_{2j+1}\) for
\(P(X = 1 \mid A = (2j{+}1)\varepsilon)\) --- Alice's, odd subscripts,
the subscript in each case giving the angle in multiples of
\(\varepsilon\).

Now run the chain in the Extreme Non-classical idealization. Each
adjacent link exhibits a perfect correlation, so the Lemma's part (a),
applied to the probabilities conditional on that link's settings, gives
\(q_{2j+1} = p_{2j}\) and \(p_{2j+2} = q_{2j+1}\) --- with No-Signaling
converting each two-setting marginal into the corresponding one-setting
marginal, which is what allows the equalities to be \emph{chained}
across experiments that share only one setting. The closing link
exhibits a perfect anti-correlation, so part (b) gives
\(q_{N-1} = 1 - p_{0}\) (with \(N\) even, so that \(N - 1\) is one of
Alice's subscripts). Chaining,

\[p_{0} = q_{1} = p_{2} = q_{3} = \cdots = q_{N-1} = 1 - p_{0},\]

and we have proved what deserves to be displayed:

\textbf{The Sorites Theorem.} If
\(p_{0} = q_{1} = p_{2} = \cdots = q_{N-1} = 1 - p_{0}\), then all of
these probabilities equal \(\tfrac{1}{2}\).

\emph{Proof.} The displayed equalities give \(p_{0} = 1 - p_{0}\).
\(\square\)

So Weak Autonomy and No-Signaling, plus the Extreme Non-classical
correlations, force every marginal in the chain to be exactly
\(\tfrac{1}{2}\) --- which is, of course, just what quantum mechanics
predicts for the singlet. The point of the derivation is not the number;
it is what the derivation reveals about the \emph{strength} of the
quantum predictions. Perfect correlations are maximally strong ---
probability \(1\) is as high as probability goes. Marginals of
\(\tfrac{1}{2}\) are maximally \emph{weak} --- every departure from
\(\tfrac{1}{2}\) would be more informative. The Sorites Theorem says
these two features are not independent: in a chained family, maximal
correlation strength \emph{forces} minimal marginal strength. One cannot
sharpen the marginals while keeping the correlations. In this
conditional sense the quantum Sorites probabilities are maximally
strong: they leave no room --- none whatever --- for improvement. That
observation is about surface probabilities. The Stronger Theorem is what
happens when a hidden variable tries to find room anyway.

\section{The Argument at the Hidden
Level}

Let \(\Lambda\) be any logical partition of propositions --- the values
of a hidden variable, under any interpretation whatsoever --- and let
\(\lambda\) denote an arbitrary member. Nothing in what follows assumes
that \(\lambda\) describes common causes fixed at the source, or
properties of the particles, or anything else specific; \(\lambda\) is
whatever additional fact a hidden-variable theorist thinks there is.
This interpretive neutrality is bought cheaply here and cashed
expensively in Part 2.

Begin with the perfect correlations of the Extreme Non-classical
idealization: for adjacent settings, \(P(X = Y \mid A = a, B = b) = 1\).
Expand this probability as an average of probabilities conditional on
the \(\lambda\)'s. An average of quantities, each at most \(1\), can
equal \(1\) only if every quantity receiving positive weight equals
\(1\). Therefore, for every \(\lambda\) with
\(P(\lambda \mid A = a, B = b) > 0\):

\[P(X = Y \mid A = a, B = b, \lambda) = 1.\]

The perfect correlations survive conditioning on \(\lambda\) --- this is
the crucial inheritance, and it required no assumption about \(\lambda\)
at all beyond the positivity of the relevant conditional probabilities.
To keep the bookkeeping clean across the chain we need that positivity
to hold uniformly:

\textbf{Weak Hidden Autonomy (Weak HA).} For all
\(\lambda \in \Lambda\): if \(P(A = a, B = b, \lambda) > 0\) for some
\((a, b) \in S\), then \(P(A = a, B = b, \lambda) > 0\) for all
\((a, b) \in S\).

Weak HA says that no value of \(\lambda\) conspires with the settings: a
\(\lambda\) that can co-occur with one of the chained setting pairs can
co-occur with any of them. It is far weaker than the usual ``hidden
autonomy'' or ``free choice'' assumption (section 6), which demands that
\(\lambda\) be statistically \emph{independent} of the settings; Weak HA
tolerates arbitrary correlations between \(\lambda\) and the settings,
short of strict exclusion.

Given Weak HA, the Colbeck--Renner Lemma applies conditional on each
viable \(\lambda\): for adjacent settings,

\[P(X = 1 \mid A = a, B = b, \lambda) = P(Y = 1 \mid A = a, B = b, \lambda).\]

To chain these equalities we need to strip the irrelevant setting out of
each conditional marginal --- the \(\lambda\)-level analogue of
No-Signaling. That assumption has an established name:

\textbf{Parameter Independence (PI).} For all \((a, b) \in S\) and all
viable \(\lambda\):
\(P(X = 1 \mid A = a, B = b, \lambda) = P(X = 1 \mid A = a, \lambda)\)
and
\(P(Y = 1 \mid A = a, B = b, \lambda) = P(Y = 1 \mid B = b, \lambda)\).

In words: once \(\lambda\) is fixed, the setting on the far wing is
probabilistically irrelevant to the outcome on this wing. PI is
\emph{not} implied by No-Signaling: the surface condition is PI's
average over \(\lambda\), and an average can be setting-independent
while its components are not --- the loophole that Part 2 will drive a
truck through. Finally, the assumption that gives hidden variables their
point:

\textbf{Improved Predictions.} Conditioning on \(\lambda\) improves the
predictions of quantum mechanics non-trivially: for at least one viable
\(\lambda\) and one chained experiment, some probability conditional on
\(\lambda\) differs from the corresponding surface probability.

Most, if not all, hidden-variable theorists want Improved Predictions;
variables that leave every prediction untouched are trivially available
(let \(\lambda\) record the color of Bob's tie) and explain nothing. Now
define \(p_{2j}(\lambda) = P(Y = 1 \mid B = 2j\varepsilon, \lambda)\)
and
\(q_{2j+1}(\lambda) = P(X = 1 \mid A = (2j{+}1)\varepsilon, \lambda)\),
exactly as before but conditional on \(\lambda\). By the inherited
perfect correlations, the Lemma, and PI --- PI now doing for the
\(\lambda\)-conditional marginals what No-Signaling did for the surface
marginals --- every viable \(\lambda\) satisfies the full chain of
equalities:

\[p_{0}(\lambda) = q_{1}(\lambda) = p_{2}(\lambda) = \cdots = q_{N-1}(\lambda) = 1 - p_{0}(\lambda),\]

and the Sorites Theorem forces every one of these conditional
probabilities to equal \(\tfrac{1}{2}\) --- the surface values.
Conditioning on \(\lambda\) improves nothing, for any viable
\(\lambda\), contradicting Improved Predictions.

That is the skeleton of the argument, laid bare in the idealization:
perfect correlations survive conditioning; the Lemma converts them into
equalities of conditional marginals; PI --- and PI alone --- lets the
equalities be chained across experiments; the Sorites Theorem then
crushes every conditional marginal to \(\tfrac{1}{2}\). Notice what is
absent: Outcome Independence was never mentioned. The Colbeck--Renner
Lemma did the work OI does in Bell's argument, and it did that work
using only the axioms of probability plus the extremal strength of the
Sorites correlations. But the skeleton is not yet a theorem, because the
Extreme Non-classical probabilities are not the true ones, and they are
not to be smuggled in as a premise. Section 6 supplies the theorem ---
same skeleton, true quantum probabilities, and, perhaps surprisingly, no
limit argument.

\section{The Stronger Theorem, Made
Rigorous}

The gap to be closed is plain. The true quantum mismatch probability at
adjacent settings is \(\sin^{2}(\varepsilon/2)\) --- quadratically
small, but not zero --- and probability \(1 - \sin^{2}(\varepsilon/2)\)
is not probability \(1\): near-perfect correlations do not survive
conditioning on \(\lambda\) the way perfect ones do, so the argument of
section 5 does not literally apply to the world. The 2014 version of
this paper closed the gap by appeal to the limit \(N \to \infty\),
waving toward the supplement of Colbeck and Renner (2011) for the
analysis. Here is a self-contained closure --- and, it turns out, one
that involves no limit at all: the contradiction is reached at a
\emph{finite} chain length, computable from the hidden-variable
theorist's own advertised margin of improvement.

Two auxiliary premises need to be strengthened, modestly, from their
section-5 forms. First, replace Weak HA by the standard autonomy (``free
choice'') assumption --- the one that every discussion of Bell's theorem
grants unless it is explicitly trafficking in conspiracies:

\textbf{Hidden Autonomy (HA).} For all \(\lambda\) and all
\((a, b) \in S\): \(P(\lambda \mid A = a, B = b) = P(\lambda)\).

(The strengthening is needed because the finite-\(N\) argument adds
error bounds across the links of the chain, and the addition requires
the \(\lambda\)-weights to be the same from link to link --- which is
exactly what HA provides and Weak HA does not.) Second, make Improved
Predictions say what a hidden-variable theorist actually wants it to say
--- that the improvement is a fact about nature, not an artifact of
coarse experiments that evaporates as the experimental family is
refined:

\textbf{Robust Improved Predictions.} There is a fixed margin
\(\eta > 0\) and a fixed weight \(w > 0\) such that, for every
sufficiently fine chain, a set of \(\lambda\)'s of total probability at
least \(w\) has some conditional chain marginal differing from
\(\tfrac{1}{2}\) by at least \(\eta\).

A \emph{deterministic} hidden variable --- one that fixes the outcomes,
so that every \(p(\lambda)\) and \(q(\lambda)\) is \(0\) or \(1\) ---
satisfies Robust Improved Predictions automatically, with
\(\eta = \tfrac{1}{2}\) and \(w = 1\). Now:

\textbf{The Stronger Theorem.} The following four assumptions are
jointly inconsistent: (1) the true quantum probabilities for the chained
singlet experiments, for all sufficiently fine chains; (2) Hidden
Autonomy; (3) Parameter Independence; (4) Robust Improved Predictions.

\emph{Proof.} Fix a chain of \(N\) links. Conditional on any
\(\lambda\), apply the Lemma's error parts to each link in turn ---
(a\('\)) on the solid links, (b\('\)) on the closing link --- using PI
at each step to strip the irrelevant setting exactly as in section 5;
the equalities of the idealized argument become inequalities, and
chaining them yields

\[\bigl|\, 2p_{0}(\lambda) - 1 \,\bigr| \;\leq\; \sum_{\text{links } k} \delta_{k}(\lambda),\]

where \(\delta_{k}(\lambda)\) is the \(\lambda\)-conditional probability
of a mismatch on link \(k\) (a disagreement on a solid link; an
agreement on the closing link). Average over \(\lambda\). By HA the
\(\lambda\)-weights are the same for every link, so the average passes
through the sum, and the \(\lambda\)-average of \(\delta_{k}(\lambda)\)
is the \emph{surface} mismatch probability of link \(k\), which quantum
mechanics puts at \(\sin^{2}(\varepsilon/2)\) per link. With \(N\) links
and \(\varepsilon = 180^{\circ}/N\):

\[\mathbb{E}_{\lambda}\, \bigl|\, 2p_{0}(\lambda) - 1 \,\bigr| \;\leq\; N \sin^{2}\!\left(\frac{\pi}{2N}\right) \;\leq\; \frac{\pi^{2}}{4N},\]

and the same bound holds for every other conditional marginal in the
chain. By Markov's inequality --- the elementary fact that a
non-negative quantity whose average is \(\mu\) can be as large as \(t\)
on at most a fraction \(\mu/t\) of cases --- the total probability of
the \(\lambda\)'s whose conditional marginals differ from
\(\tfrac{1}{2}\) by \(\eta\) or more is at most \(\pi^{2}/(8N\eta)\).
Now let the hidden-variable theorist name the margins \(\eta\) and \(w\)
of Robust Improved Predictions, and take any chain finer than
\(N^{*} = \pi^{2}/(8\eta w)\): the \(\lambda\)'s that deliver the
advertised improvement have total probability at most
\(\pi^{2}/(8N\eta) < w\). Contradiction. \(\square\)

Three remarks. First, \emph{no limit is taken anywhere in this proof}.
The contradiction is a strictly finite affair: for margins
\(\eta = w = 0.1\), a chain of \(N^{*} \approx 124\) links already does
it, and the only fact about large \(N\) that the proof uses is the
archimedean triviality that \(\pi^{2}/8N\) eventually falls below any
fixed positive number. Colbeck and Renner's own presentation proceeds
via a limit; the reorganization above shows the limit was a convenience,
not a load-bearing wall. The Extreme Non-classical idealization,
likewise, appears nowhere: every probability in the proof is a true
quantum probability at finite \(\varepsilon\). Second, for deterministic
hidden variables the conclusion is especially stark: with
\(\eta = \tfrac{1}{2}\), \(w = 1\), the theorem says that any
settings-independent, parameter-independent variable that determines the
outcomes is already contradicted by the quantum statistics of a
\(25\)-link chain --- and refining the chain squeezes the tolerated
fraction of determined runs toward zero. Third, the theorem's premise
(1) is pure quantum phenomenology, confirmed wherever it has been tested
(Aspect, Grangier and Roger 1982; Hensen et al.~2015); premise (2) is
free choice; premise (4) is the hidden-variable programme's own job
description. The arrow of \emph{modus tollens} has one place left to
land.\footnote{Ghirardi and Romano (2013) criticized the assumptions of
  Colbeck and Renner's information-theoretic packaging; the present
  formulation is designed to be immune to those criticisms by making
  every assumption explicit and elementary. Measure-theoretic pedantry
  --- the measurability of \(\lambda \mapsto p_{k}(\lambda)\) --- is
  assumed without comment, as it must be for the conditional
  probabilities to be defined at all.}

\section{Why the Stronger Theorem Is
Stronger}

Bell's theorem derives its contradiction from: the quantum predictions;
Hidden Autonomy; Parameter Independence; and Outcome Independence:

\textbf{OI} For all \((a, b) \in S\) and viable \(\lambda\):
\(P(X = 1, Y = 1 \mid A = a, B = b, \lambda) = P(X = 1 \mid A = a, B = b, \lambda)\, P(Y = 1 \mid A = a, B = b, \lambda)\).

The logical relationships between the two theorems' auxiliary
assumptions are strict entailments:

\begin{enumerate}
\def\labelenumi{(\alph{enumi})}
\item
  The autonomy premises match: both theorems use HA (and HA entails the
  Weak HA of section 5's scaffolding: if
  \(P(\lambda \mid A = a, B = b) = P(\lambda)\) and
  \(P(A = a, B = b, \lambda) > 0\) for some chained pair, then
  \(P(\lambda) > 0\), and Weak Autonomy plus HA give
  \(P(A = a', B = b', \lambda) > 0\) for every chained pair).
\item
  OI (with the true surface correlations and HA) entails Robust Improved
  Predictions. At adjacent chained settings the surface probability of
  joint agreement in the \(1\)-cells is close to \(\tfrac{1}{2}\), while
  the product of the surface marginals is \(\tfrac{1}{4}\): the outcomes
  are strongly correlated, by a gap of about \(\tfrac{1}{4}\) that does
  not shrink as the chain is refined. Suppose OI held and Robust
  Improved Predictions failed with margins as generous as
  \(\eta = w = \tfrac{1}{16}\): then, outside a \(\lambda\)-set of
  probability \(\tfrac{1}{16}\), every conditional marginal would lie
  within \(\tfrac{1}{16}\) of \(\tfrac{1}{2}\), so every conditional
  \emph{joint} --- a product, by OI --- would lie within about
  \(\tfrac{1}{16}\) of \(\tfrac{1}{4}\); averaging over \(\lambda\) (HA)
  would put the surface joint within roughly
  \(\tfrac{1}{16} + \tfrac{1}{16}\) of \(\tfrac{1}{4}\), contradicting
  the observed value near \(\tfrac{1}{2}\). So an OI theorist is
  \emph{committed} to Robust Improved Predictions.
\end{enumerate}

Given (a) and (b), Bell's premise set entails the Stronger Theorem's
premise set, so Bell's theorem is a corollary of the Stronger Theorem;
and since the entailment in (b) does not reverse --- Robust Improved
Predictions carries no commitment to factorized conditional joints ---
the Stronger Theorem is strictly stronger. The philosophical payoff is
the redistribution of blame. Bell's contradiction can be escaped by
rejecting OI alone --- the escape route recommended by Jarrett (1984)
and Butterfield (1989), and defended as the peaceful-coexistence option
by Ghirardi (2010), on the ground that OI-violation is mere correlation
while PI-violation smells of action at a distance. The Stronger Theorem
closes that route: its contradiction does not use OI, so rejecting OI
does not touch it. A hidden-variable theorist who keeps (Robust)
Improved Predictions must reject Weak HA/HA --- settings conspiracies, a
desperate move --- or reject Parameter Independence. The Sorites
argument thus does what Bell's argument alone could not: it makes
Parameter Independence \emph{the} obvious candidate for refutation.

\section{Why Parameter Independence Seemed
Safe}

If the argument of sections 2--7 succeeds, it convicts the least likely
suspect. Two considerations have long made Parameter Independence seem
the assumption a reasonable physicist would defend to the end.

\emph{(a) The relativistic consideration.} Suppose, as nearly everyone
does, that all physical processes take place in four-dimensional
spacetime. A violation of PI is then naturally read as a physical
process running from Bob's setting event to Alice's outcome event. In
the interesting experiments those events are spacelike separated; a
process connecting them would be superluminal; and special relativity
forbids superluminal processes --- or at least superluminal
\emph{causation}, on pain of the usual paradoxes of temporal order
(which spacelike separation makes frame-relative). A violation of OI, by
contrast, can be shrugged off as correlation without causation. This
asymmetry is exactly why the literature preferred sacrificing OI --- and
note, for later, that the argument's engine is the italicized
supposition: that spacetime is where every physical process must live.

\emph{(b) The inductive consideration.} Average Parameter Independence
over \(\lambda\) (with weights given by HA) and No-Signaling falls out:
if every \(\lambda\)-conditional marginal is independent of the distant
setting, so is their average. No-Signaling is not merely well confirmed;
it is built into the empirical adequacy of quantum mechanics itself and
tested wherever entangled states are manipulated. So PI has a true
consequence of great generality. The support this lends PI is inductive,
not deductive --- a true consequence never proves its premise, and
section 5's loophole (an average may wash out what its components
contain) is precisely the gap between the two. But abduction has force:
the simplest explanation of statistical locality at every setting, one
might think, is probabilistic locality at every \(\lambda\).

Part 2 now shows that both considerations fail together, and
instructively, in a concrete dynamics: (a) fails because the process
that violates PI does not run its course in spacetime at all; (b) fails
because the very model that violates PI \emph{entails} No-Signaling ---
the average washes out the violation not by accident but by a symmetry
of the dynamics.

\section{Part 2: How the Random-Matrix Framework Answers the
Argument}

\section{The Random-Matrix
Framework}

The framework, developed by Kryukov (2020; 2025; 2026; 2026a; 2026b),
can be summarized in four claims; the reader is referred to those papers
for the mathematics.

First, the arena of quantum dynamics is the space of states ---
projective Hilbert space with the Fubini--Study metric --- and
\emph{classical} reality corresponds to a special submanifold of it: the
states in which particles have definite (detector-resolution) positions
and momenta are localized wave-packets, and these form a flat classical
submanifold inside the curved state space. States on the submanifold
behave classically: constraining the Schrödinger dynamics to it yields
exactly Newtonian mechanics (Kryukov 2020; 2026).

Second, measurement is a dynamical process on the state space. When a
microscopic system in a superposition interacts with a macroscopic
device, the interaction adds to the Hamiltonian a rapidly fluctuating
random component --- modeled as a random matrix drawn afresh, at each
instant, from the Gaussian Unitary Ensemble (GUE) --- and the system's
state consequently performs a random walk in the projective space:
stochastic in its increments, but unitary at every step (Kryukov 2025;
2026a).

Third, the walk terminates on the classical submanifold: the measurement
is complete when the state reaches one of the detector-defined outcome
classes --- for a two-outcome spin measurement, one of the two localized
deflection classes. The probability of landing in a given class equals
that class's initial squared amplitude: the Born rule, derived rather
than postulated, as the exit-probability structure of a drift-free walk
(Kryukov 2025; 2026b). In the representation most useful below, the
squared amplitude \(m\) of one outcome class performs a drift-free
random walk on \([0,1]\) --- a gambler's ruin --- absorbed at the
endpoints, and optional stopping gives
\(P(\text{absorb at } 1) = m_{0}\) exactly.

Fourth, for an entangled pair the walking object is the pair's
\emph{joint} state --- a single point in the joint projective space,
since an entangled state is precisely one that does not factor into a
state for each particle. A measurement on one member drives a walk of
this single point, whose absorbing classes are the joint outcome
classes; for the singlet measured along \(b\) on Bob's side, the initial
amplitudes sit entirely on the two anti-correlated classes (``Bob up
along \(b\), Alice down along \(b\)'' and the reverse), each with mass
\(\tfrac{1}{2}\), so the walk lands in one of those two --- fixing
\emph{both} particles' spins along \(b\) in a single landing --- and
never in the other two, which carry no amplitude (Kryukov 2026a; 2026b).

The framework's credentials on the phenomena of Part 1 are the standard
quantum ones, since it reproduces the Born rule: perfect chained
correlations in the \(\varepsilon \to 0\) limit, marginals of
\(\tfrac{1}{2}\), the Sorites phenomenon in full. The question is which
premise of the Stronger Theorem it rejects. It had better reject one:
the model adds structure beyond the quantum state --- the realized
random matrices --- and conditioning on that structure determines
outcomes exactly, so (Robust) Improved Predictions holds \emph{in
principle}. Weak HA and HA are unobjectionable in the model (the random
stream is independent of anyone's choice of settings). The quantum
probabilities are its own predictions. That leaves one candidate, and
the next sections show that the model does reject it, how, and with what
compensations.

\section{The Hidden Variable Reinterpreted: Streams and
Seeds}

The theorems of Part 1 place no interpretation on \(\lambda\). The
literature's default reading --- \(\lambda\) as common causes stamped
into the particles at the source --- is optional, and the random-matrix
framework invites a different one.

Let \(\lambda\) have two independent parts,
\(\lambda = (\lambda_{a}, \lambda_{b})\), one for each laboratory.
\(\lambda_{b}\) is the realized random stream of Bob's measurement
interaction: the full list of random-matrix increments that his
device--environment complex will feed into the measurement walk ---
fixed, on a given run, by the microstate of his apparatus and its
surroundings. The right computational analogy: \(\lambda_{b}\) is like
the list of pseudo-random numbers stored on a computer that uses them to
simulate random phenomena; Bob's choice of measurement direction \(b\)
is like a \emph{random seed}. The same stored list, consumed under a
different seed, produces an entirely different-looking simulation.
Likewise \(\lambda_{a}\) for Alice. Together, \(\lambda_{a}\), Alice's
setting \(a\), and \emph{the state of the system presented to her
device} determine the outcome of her measurement; and likewise for Bob.
Given the streams, nothing is left to chance: the model is deterministic
conditional on \(\lambda\), and all quantum randomness is ignorance of
the streams.

The clause italicized above is where locality will crack, so let us make
the dependence explicit. Suppose Bob's measurement is performed first
--- the story can be told in any frame; pick one and tell it there. For
Bob, the system is the singlet pair, and his measurement drives the
joint walk of section 9: given his angle \(b\), the pair's state
collapses onto one of the two anti-correlated classes --- outcome
\(0\text{–}1\) or \(1\text{–}0\), in the notation ``Bob's bit--Alice's
bit'' --- and \emph{which} one is determined by \(\lambda_{b}\)
(together with \(b\)). That landing settles the state of the system that
Alice's device will subsequently measure: her particle now has a
definite spin along \(b\), opposite to Bob's. Her own walk, driven by
\(\lambda_{a}\) and \(a\) on the state Bob's measurement left her, then
determines her record. So Alice's outcome is a function
\(X(a, \lambda_{a}; \text{state presented})\), and the state presented
is a function of \(b\) and \(\lambda_{b}\). Bob's setting has entered
Alice's outcome --- through the front door.

\section{How the Sorites Argument Is
Blocked}

That is the qualitative shape; here is the quantitative heart. The key
claim is a sensitivity claim: \emph{at fixed \(\lambda_{b}\), the
landing of Bob's walk depends on \(b\), and near critical angles it
depends on \(b\) with unlimited sensitivity.} Random walks magnify small
differences in their driving terms into gross differences in where they
end up; two walks driven by almost --- but not exactly --- the same
increments part company eventually, and first-passage outcomes are
decided by margins that can be arbitrarily fine. A change of setting
from \(b\) to \(b + \Delta b\) perturbs every increment of the walk (the
stream is fixed; the interaction it drives is oriented differently), and
there are angles at which an arbitrarily small \(\Delta b\) flips the
landing from \(0\text{–}1\) to \(1\text{–}0\).

This claim can be checked, and Figure 3 reports a simulation designed to
check it in the most transparent representation of the walk --- the
gambler's ruin on the class mass \(m\), which the framework itself
singles out as the coordinates in which the Born rule is visible. On
each run, the mass of the class \(0\text{–}1\) starts at
\(m_{0} = \tfrac{1}{2}\) (the singlet's Born weight) and steps by
\(m \mapsto m + \epsilon\, \xi_{k}(b) \sqrt{m(1 - m)}\), absorbed at
\(0\) and \(1\); the increments \(\xi_{k}(b)\) are read off from a fixed
random stream --- the same stored numbers --- consumed through the
geometry of the device at orientation \(b\). Two facts about this
construction are exact --- theorems about the model, not observations
about the runs. (i) \emph{Born-faithfulness.} The walk of \(m\) is a
fair game: at every step, the expected value of \(m\) after the step
equals its value before (the increments have mean zero, whatever \(b\)
is). A fair game stopped when it first reaches \(0\) or \(1\) is still
fair, so the probability of ending at \(1\) must equal the starting
level \(m_{0}\) --- that is the whole argument, and it delivers the Born
rule exactly, for every setting. (ii) \emph{No-Signaling.} At any fixed
\(b\), the increments the stream delivers are standard Gaussian random
numbers, and --- here the isotropy of the random-matrix ensemble does
the work --- their \emph{statistics} are the same at every \(b\): no
direction in the state space is special, so no setting is special. Since
the statistics of the increments do not depend on \(b\), neither do the
statistics of the outcomes: Bob's marginal is \(\tfrac{1}{2}\) whatever
he measures. What depends on \(b\) is not the statistics but the
\emph{particular numbers}: at fixed stream, changing \(b\) smoothly
reshuffles which increments the walk actually receives. How finely the
reshuffling varies with \(b\) reflects the complexity of the measuring
device --- for a device with \(n\) relevant microscopic levels, the
increment at each step traces a smooth curve with roughly \(n\)
oscillations as \(b\) sweeps from \(0^{\circ}\) to \(180^{\circ}\). The
figure displays \(n = 60\); for a real magnet, \(n\) is astronomical.

\begin{figure}[t]
\centering
\includegraphics[width=\linewidth]{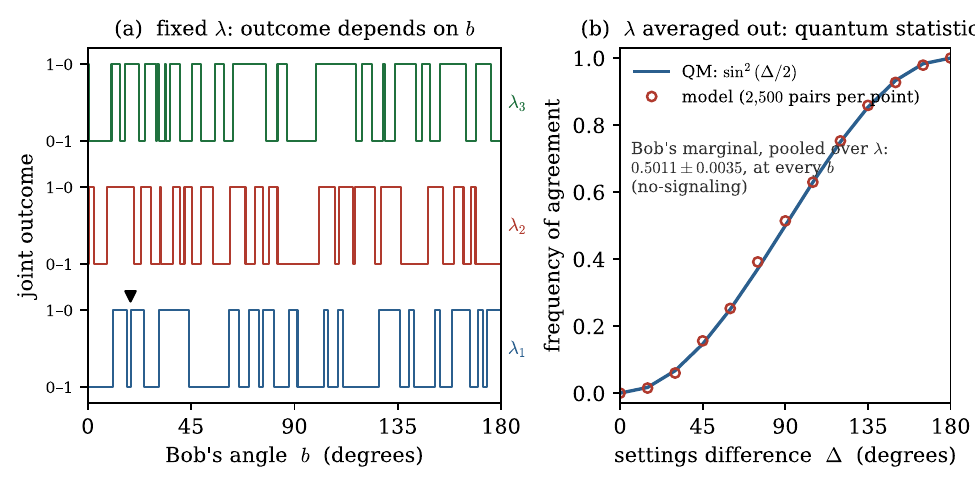}
\caption{The simulation. (a) Fixed stream \(\lambda\), swept setting \(b\): the joint outcome (\(1\text{–}0\) = Bob up--Alice down, \(0\text{–}1\) = the reverse) holds steady over stretches of \(b\) and then jumps --- roughly thirty jumps as \(b\) sweeps from \(0^{\circ}\) to \(180^{\circ}\) --- and the jump pattern differs erratically from stream to stream (\(\lambda_{1}, \lambda_{2}, \lambda_{3}\)). Zooming in on the marked jump, again and again, always finds the same thing: a \(b\) below which the outcome is one way and a \(b\) above which it is the other, boxed here inside an interval of \(1.5 \times 10^{-11}\) degrees. A change of Bob's setting smaller than a hundred-billionth of a degree reverses both records: Parameter Independence is violated, deterministically and violently. (b) The same model with the stream averaged out: Bob's marginal is \(\tfrac{1}{2}\) at every \(b\) (pooled estimate \(0.5011 \pm 0.0035\); exactly \(\tfrac{1}{2}\) in law), and the frequency with which the two wings' records agree tracks the singlet law \(\sin^{2}(\Delta/2)\) across all setting differences (\(2{,}500\) pairs per point; largest deviation \(0.021\), within Monte Carlo error). No-Signaling and the Born rule hold; nothing at the surface betrays the sensitivity underneath.}
\end{figure}

The simulated facts, then: at fixed \(\lambda_{b}\), Bob's outcome as a
function of \(b\) is a step function with many jumps, whose locations
are unpredictable from one stream to the next; across a jump, a setting
change of less than \(10^{-10}\) degrees reverses the outcome --- and
with it, in the same landing, the state of Alice's particle, and hence
(with high probability, once she measures nearby angles) her record.
Meanwhile every \(\lambda\)-averaged quantity is exactly quantum.

Now hold this model up against the Stronger Theorem, premise by premise.
The quantum probabilities: satisfied --- by construction, fact (i). Weak
HA and HA: satisfied --- the streams are set by device microstates,
independent of the settings anyone chooses. Improved Predictions, indeed
Robust Improved Predictions: satisfied --- conditional on \(\lambda\),
every outcome probability is \(0\) or \(1\), as far from
\(\tfrac{1}{2}\) as probabilities go. Outcome Independence: satisfied,
and trivially so --- probabilities of \(0\) and \(1\) factorize
automatically, so conditioning on the other wing's outcome changes
nothing. (This is the Causal Markov Condition keeping its word: the pair
of streams \((\lambda_{a}, \lambda_{b})\), together with the settings
and the source's preparation, is a \emph{complete} common cause of the
two records, and outcomes screened off by a complete common cause are
independent. The model thus occupies the quadrant the
peaceful-coexistence literature treated as uninhabitable --- OI true, PI
false --- and section 3's moral stands: it was never outcome information
that the distant wing needed.) Parameter Independence: \emph{violated}
--- \(P(Y = 1 \mid B = b, \lambda)\) and
\(P(Y = 1 \mid B = b + \Delta b, \lambda)\) differ by \(1\) across a
jump; and \(P(X = 1 \mid A = a, B = b, \lambda)\) inherits the
\(b\)-dependence through the state Bob's landing hands to Alice. The
inconsistency proved in Part 1 is real, and the model resolves it in
exactly the place the Sorites argument said someone must: the chain of
equalities \(p_{0}(\lambda) = q_{1}(\lambda) = \cdots\) never gets
started, because the equalities that PI was needed to underwrite are
false. Conditional on \(\lambda\), the marginals are not even
well-defined functions of one setting alone --- \(q_{1}(\lambda)\)
depends on what Bob's setting was --- so the Sorites Theorem has nothing
to bite on. The \(\lambda\)-conditional probabilities escape the forced
march to \(\tfrac{1}{2}\), and the model's hidden variable improves the
quantum predictions \emph{in principle} --- all the way to determinism,
which is incidentally why Outcome Independence holds for free --- while
remaining, as we are about to see, undetectable \emph{in practice}.

\section{No-Signaling Without Parameter
Independence}

The model thus discharges the two considerations of section 8, and it
pays to see exactly how.

\emph{The relativistic consideration, discharged.} In the framework, the
process that carries the joint state from the singlet to a product class
is a walk \emph{in the state space of the pair}, not a disturbance
propagating through spacetime. The pair's state is one point; the
landing is one event of the state-space dynamics; the ``influence'' of
\(b\) on Alice's particle is the utterly ordinary dependence of a
dynamical process on its own boundary conditions --- Bob's device
orientation is part of what shapes the walk that the one joint state
undergoes. Nothing traverses the space between the laboratories: on the
classical submanifold --- the arena of spacetime processes, where
relativistic causal structure lives --- nothing moves at all until the
records form, and the records form locally in each laboratory. In
Kryukov's words, ``there is no motion in classical space and no signal
or causal influence propagates between spatially separated particles
during measurement'' (2026a). The supposition that powered the
relativistic worry --- that a PI violation must be a spacetime process
--- is exactly the supposition the framework denies. What relativity
actually forbids, on any careful reading, is superluminal
\emph{signaling}; and that brings us to the second
consideration.\footnote{A note on terminology, to forestall an apparent
  conflict. Kryukov (2026a) writes that the framework ``preserves
  parameter independence: measurement settings on one side do not affect
  the marginal outcome statistics on the other.'' The quantity preserved
  there is the \(\lambda\)-\emph{averaged}, statistical independence of
  the marginals --- No-Signaling, in the standard terminology adopted
  throughout this paper (van Fraassen's 1982 Surface Locality).
  Jarrett--Shimony Parameter Independence, the
  \(\lambda\)-\emph{conditional} condition used in Bell-type theorems
  and throughout this paper, is a different and stronger claim, and it
  is the one the framework violates. Both statements are true of the
  model; they answer different questions. The precedent is instructive:
  Bohm's (1952) theory likewise violates conditional Parameter
  Independence --- the distant setting enters the guidance of the local
  particle --- while its equilibrium statistics are signal-free.}

\emph{The inductive consideration, discharged.} No-Signaling is true in
the model --- not approximately, not contingently, but as a theorem
(fact (ii) of section 11): the distribution of the fixed stream's
increments is the same at every setting, so Bob's choice of \(b\) leaves
Alice's outcome statistics untouched, and hers leave his. The inductive
argument for PI --- ``No-Signaling is true, and PI is its best
explanation'' --- is thereby defused by counterexample: here is a model
in which No-Signaling has a \emph{different} explanation, a symmetry
(the isotropy of the random-matrix ensemble) rather than a
\(\lambda\)-level independence. The average washes out the violation
because the violation has no preferred direction, not because there is
nothing to wash out.

And why can Bob not exploit his very real influence to send a message?
Two locked doors stand in the way, and both are visible in Figure 3(a).
First, to steer Alice's outcome Bob would need to know \emph{which side
of a jump} his setting sits on, and the jump locations are functions of
\(\lambda_{b}\) --- the microstate of his own device and environment
down to the random stream it will generate --- which he can neither read
nor prepare. The sensitivity that makes his influence violent (flips
across \(10^{-10}\) degrees) is precisely what makes it unusable: an
influence whose direction depends chaotically on unknowable initial
conditions cannot carry a chosen bit. Second, even the \emph{fact} of
correlation is invisible to Alice alone: her local statistics are
\(\tfrac{1}{2}\)--\(\tfrac{1}{2}\) whatever Bob does, and the
correlation between the wings appears only when the two records are
brought together through an ordinary, light-limited channel. Bob's
influence on Alice's outcome is like a hand that cannot help trembling:
strong enough to knock the coin from heads to tails, far too unsteady to
place it. So the model sits exactly where quantum mechanics has always
sat --- outcomes correlated, signals silent --- but now with a mechanism
underneath: Parameter Independence false, Outcome Independence true,
No-Signaling true, and no tension anywhere, because the violation is a
fact about the realization and the two preserved conditions are facts
about the law.

\section{Conclusion}

Part 1 argued that the quantum Sorites phenomenon --- the chained,
in-the-limit-perfect correlations of the singlet state --- supports a
theorem strictly stronger than Bell's: a contradiction between the
quantum predictions, a weak autonomy condition, Parameter Independence,
and the assumption that hidden variables improve the quantum
probabilities; Outcome Independence is not a premise, so rejecting it is
not an escape. Made rigorous (section 6) --- at finite chain length,
with true quantum probabilities and no limit argument --- the theorem
says: any settings-independent hidden variable that robustly improves
the quantum predictions must violate Parameter Independence. The theorem
thereby does what Bell's theorem famously could not --- it picks out a
unique culprit among the locality conditions.

Part 2 argued that the culprit can be convicted without damage to
physics. In the random-matrix collapse dynamics, the hidden variable is
the random stream that drives the measurement walk --- stored
randomness, with settings as seeds --- and Parameter Independence fails
exactly as the theorem demands: conditional on the stream, the distant
setting helps determine the local outcome, with a sensitivity
(demonstrated here down to \(10^{-10}\) degrees of magnet orientation)
that grows with the complexity of the device. The failure carries no
relativistic sting, because the process that implements it is a walk in
state space, not a signal in spacetime; and it is empirically invisible
in isolation, because the model entails No-Signaling as a symmetry of
its driving ensemble --- the quantum statistics, Born rule and all,
emerge exactly upon averaging over the stream. The Sorites argument is
blocked at the only joint the model leaves open: the chain of marginal
equalities that Parameter Independence was needed to license never
forms, so the conditional probabilities are free to be deterministic ---
different from the quantum probabilities in principle, though not in
practice, since the stream that would reveal the difference is the one
thing no experimenter can hold fixed.

The 2014 version of this paper ended by saying that hidden-variable
theorists may need to rethink the claim that rejecting Outcome
Independence is the way to keep quantum mechanics at peace with
relativity. The present version can end more constructively: there is at
least one worked-out dynamics of measurement in which the peace is kept
on the opposite terms --- Parameter Independence surrendered, Outcome
Independence and No-Signaling preserved, and the surrender explained by
the two ideas the framework is built on: that the dynamics of
measurement runs in the space of states rather than in spacetime, and
that its randomness is stored in the measuring device rather than
stamped on the particles. Whether that framework is true is a question
for its developing experimental programme. That it is \emph{possible} is
enough to complete the argument of this paper: the strongest locality
condition in the Bell literature is the one that has to go, and there is
a principled, relativity-respecting way to let it go.

\subsection{Appendix: Why GHZ Does Not Support the Stronger
Theorem}

It is natural to ask whether the perfect correlations of the
Greenberger--Horne--Zeilinger experiment (Greenberger, Horne and
Zeilinger 1989) could power the Stronger Theorem without any chaining.
They cannot, and seeing why illuminates what the Sorites structure
contributes.

In the GHZ setup, three particles fly to three devices \(A\), \(B\),
\(C\), each with two settings (\(1\) or \(2\)) and outcomes recorded as
\(X, Y, Z \in \{-1, +1\}\). Quantum mechanics predicts perfect
correlations for four joint settings: \(Y X Z = +1\) with certainty at
the settings \(1\text{-}1\text{-}2\), \(1\text{-}2\text{-}1\), and
\(2\text{-}1\text{-}1\); and \(Y X Z = -1\) with certainty at
\(2\text{-}2\text{-}2\). Weak Autonomy and No-Signaling apply here as
before. But no Sorites chain is present, and the marginals are not
forced to \(\tfrac{1}{2}\) by these correlations. Proof by example:
assign \(P(X = 1 \mid A = 1) = 1 = P(X = 1 \mid A = 2)\) --- a maximally
strong, setting-independent marginal for the first particle --- and the
four perfect correlations reduce to constraints on \(Y\) and \(Z\) alone
(\(YZ = +1\) at the three ``positive'' setting pairs, \(YZ = -1\) at
\(2\text{-}2\)), which the Colbeck--Renner Lemma converts into marginal
equalities that are jointly satisfiable (set the four \(Y\)- and
\(Z\)-marginals to \(\tfrac{1}{2}\)). Since the surface probabilities
can be consistently \emph{strengthened}, they are not maximal, and the
argument of section 5 --- which needed the maximality to squeeze the
\(\lambda\)-conditional marginals against the surface values --- cannot
begin. Of course the actual quantum marginals in the GHZ experiment are
\(\tfrac{1}{2}\); the point is that the four GHZ correlations do not
force them to be, whereas the Sorites chain forces its marginals to be
\(\tfrac{1}{2}\) twice over --- at the surface, and conditional on every
\(\lambda\). One can still derive the GHZ version of Bell's theorem by
assuming OI and HA in the usual way; what one cannot do is dispense with
OI. GHZ's correlations are perfect but few; the Sorites correlations are
perfect and \emph{chained}, and it is the chain that does the work.

\section*{References}

\noindent\hangindent=1.5em\hangafter=1 Aspect, A., Grangier, P., and Roger, G. (1982). Experimental realization
of Einstein-Podolsky-Rosen-Bohm Gedankenexperiment: A new violation of
Bell's inequalities. \emph{Physical Review Letters} 49, 91--94.

\noindent\hangindent=1.5em\hangafter=1 Bell, J. S. (1964). On the Einstein Podolsky Rosen paradox.
\emph{Physics} 1, 195--200.

\noindent\hangindent=1.5em\hangafter=1 Bohm, D. (1952). A suggested interpretation of the quantum theory in
terms of ``hidden'' variables, I and II. \emph{Physical Review} 85,
166--179 and 180--193.

\noindent\hangindent=1.5em\hangafter=1 Braunstein, S. L. and Caves, C. M. (1990). Wringing out better Bell
inequalities. \emph{Annals of Physics} 202, 22--56.

\noindent\hangindent=1.5em\hangafter=1 Clauser, J. F. and Horne, M. A. (1974). Experimental consequences of
objective local theories. \emph{Physical Review D} 10, 526--535.

\noindent\hangindent=1.5em\hangafter=1 Butterfield, J. (1989). A space-time approach to the Bell inequality. In
J. T. Cushing and E. McMullin (eds.), \emph{Philosophical Consequences
of Quantum Theory: Reflections on Bell's Theorem}, 114--144. Notre Dame,
IN: University of Notre Dame Press.

\noindent\hangindent=1.5em\hangafter=1 Colbeck, R. and Renner, R. (2011). No extension of quantum theory can
have improved predictive power. \emph{Nature Communications} 2, 411.

\noindent\hangindent=1.5em\hangafter=1 Forster, M. R. (1986). Bell's paradox and path analysis. In P.
Weingartner and G. Dorn (eds.), \emph{Foundations of Physics}. Vienna:
Hölder-Pichler-Tempsky.

\noindent\hangindent=1.5em\hangafter=1 Forster, M. R. (2014). How the quantum Sorites phenomenon strengthens
Bell's argument. arXiv:1403.1598.

\noindent\hangindent=1.5em\hangafter=1 van Fraassen, B. C. (1982). The Charybdis of realism: Epistemological
implications of Bell's inequality. \emph{Synthese} 52, 25--38.

\noindent\hangindent=1.5em\hangafter=1 Ghirardi, G. (2010). Does quantum nonlocality irremediably conflict with
special relativity? \emph{Foundations of Physics} 40, 1379--1395.

\noindent\hangindent=1.5em\hangafter=1 Ghirardi, G. C. and Romano, R. (2013). About possible extensions of
quantum theory. \emph{Foundations of Physics} 43, 881--894.

\noindent\hangindent=1.5em\hangafter=1 Graßhoff, G., Portmann, S., and Wüthrich, A. (2005). Minimal assumption
derivation of a Bell-type inequality. \emph{The British Journal for the
Philosophy of Science} 56, 663--680.

\noindent\hangindent=1.5em\hangafter=1 Greenberger, D. M., Horne, M. A., and Zeilinger, A. (1989). Going beyond
Bell's theorem. In M. Kafatos (ed.), \emph{Bell's Theorem, Quantum
Theory, and Conceptions of the Universe}, 73--76. Dordrecht: Kluwer.

\noindent\hangindent=1.5em\hangafter=1 Hensen, B., Bernien, H., Dréau, A. E., et al.~(2015). Loophole-free Bell
inequality violation using electron spins separated by 1.3 kilometres.
\emph{Nature} 526, 682--686.

\noindent\hangindent=1.5em\hangafter=1 Hofer-Szabó, G., Rédei, M., and Szabó, L. E. (2013). \emph{The Principle
of the Common Cause}. Cambridge: Cambridge University Press.

\noindent\hangindent=1.5em\hangafter=1 Jarrett, J. P. (1984). On the physical significance of the locality
conditions in the Bell arguments. \emph{Noûs} 18, 569--589.

\noindent\hangindent=1.5em\hangafter=1 Kryukov, A. A. (2020). Mathematics of the classical and the quantum.
\emph{Journal of Mathematical Physics} 61, 082101.

\noindent\hangindent=1.5em\hangafter=1 Kryukov, A. A. (2025). Dynamics of a particle in the double-slit
experiment with measurement. \emph{Journal of Physics A: Mathematical
and Theoretical} 58, 225302.

\noindent\hangindent=1.5em\hangafter=1 Kryukov, A. A. (2026). Emergence of classical dynamics from a random
matrix Schrödinger model. \emph{Physics Letters A} 589, 131791.

\noindent\hangindent=1.5em\hangafter=1 Kryukov, A. A. (2026a). Quantum paradoxes and the quantum-classical
transition under unitary measurement dynamics with random Hamiltonians.
arXiv:2601.17976.

\noindent\hangindent=1.5em\hangafter=1 Kryukov, A. A. (2026b). Random-matrix reduction in projective quantum
mechanics. Manuscript, submitted.

\noindent\hangindent=1.5em\hangafter=1 Shimony, A. (1984). Controllable and uncontrollable non-locality. In S.
Kamefuchi et al.~(eds.), \emph{Foundations of Quantum Mechanics in the
Light of New Technology}, 225--230. Tokyo: Physical Society of Japan.

\noindent\hangindent=1.5em\hangafter=1 Shimony, A. (1986). Events and processes in the quantum world. In R.
Penrose and C. J. Isham (eds.), \emph{Quantum Concepts in Space and
Time}, 182--203. Oxford: Clarendon Press.

\noindent\hangindent=1.5em\hangafter=1 Suppes, P. and Zanotti, M. (1976). On the determinism of hidden variable
theories with strict correlation and conditional statistical
independence of observables. In P. Suppes (ed.), \emph{Logic and
Probability in Quantum Mechanics}, 445--455. Dordrecht: Reidel.

\end{document}